\documentstyle[12pt]{article}
\begin{document}
\title{ A geometrical model\\
for \\ Mixed cyanide crystals }
\author{Serge Galam}
\date{Laboratoire des Milieux D\'{e}sordonn\'{e}s et H\'{e}t\'{e}rog\`{e}nes\footnotemark[1], 
\\ÊUniversit\'e Paris 6,
Tour 13 - Case 86, 4 place Jussieu, 75252 Paris Cedex 05, France\\
(e-mail: galam@ccr.jussieu.fr)\\$\ $\\
{\em Short title:\/} Orientational Glasses\\ 
Key words: percolation, glasses, random fields\\
{\em PA Classification Numbers:\/}
 64.60 A, 64.60 C, 64.70 P\\$\ $\\
(J. of Non-Crsytalline Solids \underline{235-237} (1998) 570-575)
}
\maketitle
\addtocounter{footnote}{1}
\footnotetext{Laboratoire associ\'{e} au CNRS (UMR n$^{\circ}$ 7603)}

\begin{abstract}

A model of diluted random field sustained by quenched volume deformations is shown to reproduce
puzzling physical features found in $X(CN)_{x}Y_{1-x}$ mixed cyanide crystals. $X$ is an alkali metal 
($K$, $Na$ or $Rb$) and $Y$ 
is a spherical halogen ion ($Br$, $Cl$ or $I$).
Critical thresholds $x_c$ at which associated first order ferroelastic transitions disappear 
are calculated exactly.
The diluted random field is shown to compete with compressibility in making 
the transition first order. Transitions are then found to remain first order down to $x_c$  
except in the case of $bromine$ dilution where they become continuous.
All the results are in excellent agreement with available experimental data.

\end{abstract}
\newpage
\section{Introduction}

In recent years a great deal of experimental effort has been devoted to investigating the
existence of a possible orientational glass state in mixed cyanide crystals [1, 2].
In particular diluted cyanide mixtures $X(CN)_{x}Y_{1-x}$ where $X$ is an alkali metal 
($K$, $Na$ or $Rb$)
and $Y$ stands for a spherical halogen ion 
($Br$, $Cl$ or $I$) are known to exhibit a series of puzzling physical features.

For instance, the ferroelastic transition which occurs in pure alkali-cyanides $XCN$ [3] is
found to disappear at a cyanide concentration threshold $x_c$ which is not equal to
the underlying lattice percolation threshold. The actual value of
$x_c$ varies with $X$ and $Y$ compounds. In the case $X=K$ it is $x_c=0.60$ and $x_c=0.80$
respectively for $Y=Br$ and $Y=Cl$.

In addition, below $x_c$ there exists experimental
evidence for local freezing of cyanide orientations, together with the absence of 
quadrupolar long range order. In parallel $X(CN)_{x}Y_{1-x}$ mixtures
remain cubic at low temperatures [1]. Analogies to spin-glasses and random field systems have thus
generated  much interest, in particular in elucidating the physical nature of 
the orientational glass state [1, 2].

Moreover the physics above $x_c$ is still puzzling. The transition is found to be
continuous at $x_c$ for $X(CN)_{x}Br_{1-x}$ [4] while it stays first 
order with $X(CN)_{x}Cl_{1-x}$ [5].

In this paper 
a model of diluted random field sustained by quenched volume deformations is proposed. It
is shown to reproduce
most of  puzzling physical features found in $X(CN)_{x}Y_{1-x}$ mixed cyanide crystals. 

A derivation of a microscopic calculation of critical thresholds $x_c$ is presented.
On this basis, an Hamiltonian
of a diluted compressible ferromagnetic Ising system is studied within mean field theory.
Diluted staggered fields are then included to account for the quenched volume deformations
generated by the substitution of spherical halogens to dumbbell-shaped
cyanides.

Compressibility is 
found to produce
a first order transition at cyanide concentrations larger that some threshold $x_L$. By contrast,
diluted random fields are shown to activate a first order transition only
below another cyanide concentration $x_r$.
 
Transition orders are then discussed for a large variety of mixtures. Relative amplitudes 
in $x_c$, $x_L$ and $x_r$ are instrumental in determining the transition order. 
Transitions are predicted to remain first order down to $x_c$ for all mixtures 
except in the case of dilution with $bromine$ which exhibits a continuous transition at $x_c$.
Our results fit perfectly to available experimental data.

\section{More than site dilution}

Starting from pure $XCN$, substitution of one spherical molecule $Y$ to a dumbbell-shape cyanide ion
has first the effect of diluting one rotational degree of freedom. Within a simple site dilution 
picture, associated phase diagram would be only slightly modified by dilution.
 The transition temperature would 
be a decreasing function of the ``hole" (no degree of freedom) density. Moreover it would vanish at the
lattice site percolation threshold. Here, it would mean ferroelastic transitions disappear at 
$x_c=p_c$ for all $X(CN)_{x}Y_{1-x}$ mixtures where cyanides are on a {\sf fcc} lattice
where $p_c=0.198$ [6]. However, from experiments 
$x_c\neq p_c$ and $x_c$ varies from one system to another.

Several physical arguments can explain the discrepancy between $x_c$ and $p_c$. Most of them
like for instance the existence of long range interactions would lead one to give up the 
site dilution approach. However it could be kept allowing simultaneously an effective
density of free-to-reorient cyanides. In other words, dilution would have the additional effect of 
``neutralizing" some cyanides taking them off the process of propagation of quadrupolar
long range order active in the ferroelastic transition at stake. 

The above picture is given a physical basis by postulating the existence of an effective density 
of free-to-reorient cyanides
$x_f$, such that all ferroelastic transitions disappear at the same threshold $x_{f,c}=p_c$. It implies
in parallel a density of ``neutralized" cyanides $x_n$ which depends on the mixture itself
with $x_n=x_c-p_c$ at $x=x_c$.

At this stage we do not need to elaborate on the physics of these ``neutralized" cyanides. What
matters here is their non-participation in the propagation of long range order. 
The effect being a function 
of the mixture, we define naturally $x_n$ as,
\begin{equation}
x_n\equiv  \alpha(1-x) \:,
\end{equation}
where $\alpha$ is the average number of cyanide ``neutralized" by one $Y$ ion.

\section{A geometrical model}

To implement our description, we need a quantitative calculation of $\alpha$. 
Noticing $XCN$ in bulk is cubic only on average, a local deformation of the unit cell is expected
while substituting one spherical $Y$ ion to the dumbbell-shaped cyanide. Let us then assume
local $XY$ unit cell embedded in $XCN$, to match that of $XY$ in bulk.
Such a mechanism yields a local volume deformation of former unit cell of amplitude,
\begin{equation}
\Delta v = \frac{1}{4}( a_{KCN}^3 - a_{KX}^3 ) \:, 
\end{equation} 
(4 molecules per unit cell). Each one of the $c$ $Y$-nearest neighbours 
($c=12$ on the {\sf fcc} cyanide sublattice) being affected
by these local volume deformations, the overall volume deformation affects 
on average a total volume $\Delta V=c\Delta v$.

Volume deformation $\Delta V$ in turn induces some deformation of neighbouring
unit cells. The deformable part of these cells is free volume which originates
from volume differences between a given molecule in
bulk $v_{XCN}$ and actual molecule itself $v_{o}$. We thus have a free volume per molecule,
\begin{equation}
v_f=v_{XCN}- v_{o}=\frac{a_{XCN}^3}{4}-(v_X+v_{CN})\:,
\end{equation}
where $a_{XCN}$ is the pure $XCN$ lattice constant, 
$v_X$ and $v_{CN}$ are ion volumes [7].

Dumbell-shaped $CN$ reorientations are directly coupled to the unit cell shape via steric 
hindrance mechanisms. The existence 
of a finite number of equivalent orientations results from the difference in symmetries between 
respectively the molecule and the lattice. Symmetry being lower for the molecule,
any volume deformation of a cyanide cage (increase or decrease) lowers the corresponding symmetry 
which in turn disturbs the regular dynamics of cyanide 
reorientations. More precisely the number of accessible orientations is lowered. Here we are
assuming that a cyanide whose cage is deformed (increased or decreased) does not participate in the 
propagation of long range order at stake in the ferroelastic transition. 
Accordingly one $Y$ substitution will affect on average, 
\begin{equation}
\alpha = c \frac{|\Delta v|}{v_f}\:,
\end{equation} 
unit cells by deforming their respective free volumes. 
Along these lines, the effective cyanide
density of free-to-reorient cyanides is,
\begin{equation}
x_{f}= x-\alpha(1-x) \:.
\end{equation}
The equality $x_{f,c}=p_c$ yields the critical threshold 
$x_c$,
\begin{equation}
x_c = \frac{p_c+\alpha}{1+\alpha}\:.
\end{equation}
Below $x_c$ a region of randomly oriented
ferroelastic domains with no static phase transition is obtained.
These domains shrink with increasing dilution to disappear eventually when $x_{f}=0$
at a new threshold [8],
\begin{equation}
x_{d}=\frac{\alpha}{1+\alpha}\:.
\end{equation}
Comparing Eqs. (6, 7) shows it is the non-zero value of $p_{c}$ which produces the randomly 
oriented ferroelastic domains. Domains are thus created by 
site statistical fluctuations.
Shear torque experiments [9], as well as diffraction experiments [10] suggested a very 
similar phase diagram.

Without a fitting parameter direct microscopic calculation
of $x_c$ and $x_{d}$ is readily performed using cristallographic data [11, 12].
The cyanide ion volume is 
$v_{CN}=34.47$\AA$^3$ [13].
The results are obtained for various mixtures (see the Table) including systems for which no 
experimental data are available.

\section{Dilute random fields}

Within the simplest model, cyanide orientations can be represented by ferromagnetic Ising 
spins $\{S_i\}$.
A random site
variable $\epsilon_{i}$ is then introduced to account for site dilution. It is 
one if site $i$ is occupied by a cyanide and zero otherwise. We have $\{\epsilon_{i}\}_{av} = x$, 
where $\{...\}_{av}$ denotes a configurational average over site disorder. 

Elastic degrees of
freedom are also introduced via volume fluctuations to account for the first order 
character of pure $XCN$ transition. Within an harmonic model they are integretated out to yield an
effective Hamiltonian,
 \begin{equation}
H_{eff} = -G \sum_{<i,j>}\epsilon_{i}\epsilon_{j}S_iS_j
-E (\sum_{<i,j>}
\epsilon_{i}\epsilon_{j}S_iS_j)^2\:,
\end{equation}
where $G$ and $E$ are constants [14].

With respect to the ``neutralization" phenomenon two mechanisms seem possible to block 
some dipoles in a $Y$ ion vicinity. It could be either
a local increase in orientational barriers or
the existence of quenched orientational random fields. Local symmetry is the major
difference between them.  Cubic symmetry is preserved in the first case and not in last.
However global cubic symmetry is preserved for both.
Available experimental data do not allow actual freezing process to be selected.

 Within a
two equivalent 
orientation model, a cyanide prevented from reorientation is indeed trapped along one direction.
On this basis, probability $p_{t}$ to have a local random field is identical to the probability of 
having a deformed cyanide cage.
From the density $x_t$ of trapped cyanide we obtain,
\begin{equation}
p_{t}= \left\{ \begin{array}{ll}
                 \frac{\alpha(1-x)}{x} & \mbox{if}\: x_{d} \leq x \leq 1 \\
                  1                    & \mbox{if}\: x < x_{d}
                 \end{array}
       \right. \:.
\end{equation}
Associated contribution to the Hamiltonian is,
\begin{equation}
H_{t}=-\sum_{i}\epsilon_{i}h_{i}S_{i}\:, 
\end{equation}
where $h_{i}$ is a quenched diluted non-symmetry breaking random field. 
 The probability distribution function is,
\begin{equation}
P(h_{i})=\frac{p_{t}}{2}[\delta (h_{i}-h)+\delta (h_{i}+h)]+(1-p_{t})\delta(h_{i})\:, 
\end{equation}
which satisfies both required symmetry conditions $P(h_{i})=P(-h_{i})$ and $\overline{ h_{i}}=0$ where 
the overline denotes configurational average over the fields. 
Adding this term to $H_{eff}$ results in full Hamiltonian $H_{T}=H_{eff}+H_{t}$ which
is rather complicated. 

A partial mean field treatment is now performed
(more details will be published elsewhere) to extract main physical features of the model.
The order parameter 
is $m=\{\overline{<\epsilon_{i}S_{i}>}\}_{av}$ where averages are done over both site and 
random field disorders.  
The double configurational averaged lattice site free energy is,
\begin{eqnarray}
{\cal F}&=&\frac{1}{2} cGm^{2} + \frac{3L}{4}m^4 \nonumber \\
        & &  -x k_{B}T \left [ \frac{p_{t}}{2}\{ln[cosh(\beta cGm
+\beta Lm^{3}+\beta h)] +
ln[cosh(\beta cGm+\beta Lm^{3}-\beta h)]\} \right. \nonumber \\
        & &  +(1-p_{t})ln[cosh(\beta cGm+\beta Lm^{3}){\bf ]}\left ]
\,\frac{\,\,\,}{\,\,\,} \,
 k_{B}Tln(2) \right.\:,
\end{eqnarray}
where $\beta\equiv\frac{1}{k_BT}$, $k_B$ is the Boltzman constant and $T$ is the temperature.
Site and RF disorders have been assumed to be uncorrelated in the averaging, i.e.,
variables
$\epsilon_{i}$ and $h_{i}$ are
statistically independent. The actual
value $h_{i} = h$ or $-h$ is independent of the value of
$\epsilon_{i}$ at a given
site $i$. Also, whether $h_{i} = \pm h$ or $0$ does not depend on site
$i$ itself.
A statistical dependence between $h_{i}$
and $\epsilon_{i}$ could only results from spatial correlations between CN sites which are anyhow
neglected in a MFT.

\section{Tricritical points}

An exhaustive study of Eq. (12) is out the scope of the present work.
At this stage it is useful to analyse two simple limit cases which are physically meaningful.
First it is the zero-steric hindrance effect case with $\alpha=0$.
It implies $x_c = p_c$, $x_d = 0$ and $p_t=0$. 

From a Landau expansion of Eq. (12) a continuous transition
occurs at the critical temperature, $k_{B}T_{c}=xcG$
under the condition of a positive quadratic coefficient,
\begin{equation}
B=-\frac{L}{cG}+\frac{1}{3x^2}\:,
\end{equation}
which results in the condition $x<  x_L$
on cyanide concentration where,
\begin{equation}
x_L\equiv  (\frac{cG}{3L})^{1/2}\:.
\end{equation}
At $x=x_L$ the continuous transition along $k_{B}T_{c}=xcG$ turns first order via a 
tricritical point ($B=0$ with a positive free energy sixth order coefficient). 

At $x=1$, $p_t=0$ (Eq. (9)) even if $\alpha \neq 0$. From experiment, pure $XCN$ exhibits 
a first order transition. 
Therefore all plastic systems must satisfy  $x_L < 1$ which sets condition $L>\frac{cG}{3}$.
Dilution is then found to weaken
the first order character of the transition.
Associated negative quartic term $B$ gets smaller in amplitude to vanish eventually at a 
tricritical point ($x=x_L$). 
There the transition is continuous with tricritical exponents. Upon further dilution the 
transition becomes second-order for $0\leq x< x_L$.

\section{Results}

We now consider the
zero-compressibility case ($L= 0$) to study random field effects ($\alphaÊ\neq 0$). 
At $x=1$ (no dilution) with $p_t$ an independent 
external parameter, the zero-compressibility free energy becomes identical to that of trimodal 
random field Ising models [15, 16, 17] which have been previously studied. 
A first order transition is found only for
$0.73<p_t\leq 1$ and for some restricted range of random field intensities 
$\sim 0.55 < \frac{h}{cG}< \sim 0.65$ [17].

These results produce an additional threshold in cyanide density around $p_t=0.73$ (Eq. (9)),
\begin{equation}
x_r\equiv \frac{\alpha}{0.73+\alpha}\:.
\end{equation} 
Only at $x<x_r$ dilution can turn
the transition first order via random fields. However condition $x_c<x_r$ must also be satisfied
since the transition itself disappears at $x_c$.
The equivalent constraint on $\alpha$ gives,
\begin{equation}
\alpha >\frac{0.73 p_c}{1-0.73-p_c}\sim 2.09\:.
\end{equation} 

 Eq. (22) gives $x_r=0.60$ and $x_r=0.83$ for respectively $K(CN)_{x}Br_{1-x}$ ($\alpha=1.12$)
 and $K(CN)_{x}Cl_{1-x}$
($\alpha=3.48$). In parallel we found $x_c=0.62$ and $x_c=0.82$ 
for $Br$ and $Cl$ dilution respectively (see the Table).
On this basis we conclude that upon dilution, while random fields can turn the
transition first order in $K(CN)_{x}Cl_{1-x}$ mixtures ($x_c<x_r$), they cannot 
do it for $K(CN)_{x}Br_{1-x}$ mixtures
($x_c>x_r$).

\section{Discussion}

It is worth noting that in the vicinity of
tricritical points 
mean field results are indeed reasonable. However, at this stage it is of importance to stress that 
some doubt exists with respect to the validity of the mean field result of a first order transition
for the random 
field Ising model. A bimodal distribution yields a tricritical point while a Gaussian distribution 
does not [15, 18, 19, 20]. 

In parallel Monte Carlo simulations
showed a tricritical point to occur in diluted ferromagnets in a staggered field at weak dilution once
next nearest interactions are introduced [21].  
An exact mapping was earlier found out between the mean field treatment of respectively a bimodal random 
field and a staggered field [22]. 

Since our results are in agreement with experiments we have to validate our mean field treatment 
of a bimodal distribution.  This is achieved introducing
a staggered symmetry among diluted random fields. It seems to indicate 
volume deformations may produce staggered random fields.

\section{Conclusion}

In summary, compressibility and staggered random field are active in 
making the transition first order. However compressibility weakens with dilution to fade out in the
vicinity of $x_c$ for all systems which means it is reasonable to assume $x_L\leq x_c$.
On the opposite staggered random fields start to be instrumental 
at dilution below $x_r$ which has an effect only when $x_r>x_c$, i.e., $\alpha >2.09$ (Eq. (16)). 
From the Table this condition is found to be always satisfied except for mixtures with $bromine$.

We can thus predict dilution with $chlorine$ and $iodine$
keeps the transition first order down to $x_c$. Only $bromine$ turns the transition continuous.
A tricritical point is thus expected for $X(CN)_{x}Br_{1-x}$ mixtures.
 At this stage our predictions reproduce
experimental results with respect to $potassium$ systems. Additional experiments on mixtures with
$sodium$ and $rubidium$ will check the validity of our model.

\subsection*{Acknowledgments}
The author would like to thank P. Doussineau, A. Levelut, and Y. Shapir for stimulating 
comments on the manuscript.

\newpage

\begin{table}
\caption{\sf Numerical values calculated for $\alpha$, $x_c$ and $x_d$ and experimental 
thresholds when known (denoted by `{\it exp:}'). Units for lengths
and volumes
are $\AA$  and $\AA^3$.
See details in the text. Error bars are within respectively $\pm 0.01$ for all data in the Table,
and $\pm 0.05$ 
for experimental thresholds.} 
\label{tbl}
\begin{tabular}{|l|l|l|l|l|l|l|l|} \hline  $XCN/ Y$ &$a_{XCN}$ & 
$a_{XY}$ &$\Delta v$ &$v_f$ &$\alpha$&$x_c$&$x_d$\\  \hline 
$KCN/ Cl$ & 6.53 &6.29 & 7.30& 25.19& 3.48& 0.82 {\it exp:} 0.80& 0.78 
{\it exp:} 0.75 \\ 
$KCN/ Br$ & 6.57 &6.60 & 2.36& 25.19& 1.12& 0.62 {\it exp:} 0.60& 0.53 
{\it exp:} 0.50 \\ 
$KCN/ KI$ & 6.53 &7.06 &18.46& 25.19& 8.79& 0.92 {\it exp:} 0.90& 0.90\\  
$NaCN/ Cl$& 5.90 &5.65 & 5.99& 12.67& 5.68& 0.88	{\it exp:}$\sim 0.80$ & 0.86 \\ 
$NaCN/ Br$& 5.90 &5.97 & 2.11& 12.67& 1.99& 0.73	& 0.67 \\    
$NaCN/ KI$ & 5.90 &6.47 &16.63& 12.67& 15.74& 0.95	& 0.94 \\ 
$RbCN/ Cl$& 6.82 &6.58 & 8.05& 30.98& 3.12& 0.80	& 0.76 \\ 
$RbCN/ Br$& 6.82 &6.85 & 1.19& 30.98& 0.46& 0.45 {\it exp:} 0.55 & 0.32 \\ [5pt] \hline
\end{tabular}
\end{table}

\end{document}